\documentclass[12pt]{article}

\usepackage{epsfig}

\catcode`@=11
\def\secteqno{\@addtoreset{equation}{section}%
\def\theequation{\thesection.\arabic{equation}}}
\catcode`@=12
\secteqno
\newcommand{\be}{\begin{equation}}
\newcommand{\ee}{\end{equation}}
\newcommand{\bea}{\begin{eqnarray}}
\newcommand{\eea}{\end{eqnarray}}
\newcommand{\bref}[1]{(\ref{#1})}
\newcommand{\nn}{\nonumber}
\newcommand{\A}{\alpha} \newcommand{\B}{\beta} \newcommand{\gam}{\gamma}
 \newcommand{\D}{\delta} 
 \newcommand{\vep}{\varepsilon}
\newcommand{\T}{\theta} 

\newcommand{\lam}{\lambda}\newcommand{\s}{\sigma}
          \newcommand{\w}{\omega}
          
\newcommand{\h}{\eta}

\newcommand{\Tb}{{\overline\theta}}

\def\6{\partial}
\def\7{\tilde}
\def\8{\hat}
\def\hmu{{\hat {\mu}}}
\def\hnu{{\hat {\nu}}}
\def\ha{\hat {a}}
\def\hb{\hat {b}}

\def\hmu{{\hat{\mu}}}
\def\hnu{{\hat{\nu}}}

\def\bt{\bar {\theta}}

\def\cL{\mathcal{L}}
\def\pa{\partial}

\def\CL{{\cal L}}
\def\CF{{\cal F}}
\def\t{\tilde}\def\CX{{\cal X}}\def\CY{{\cal Y}}\def\CB{{\cal B}}
\def\l{{\ell}}
\def\vs{\vskip 4mm}\def\={{\;=\;}}\def\+{{\;+\;}}
\def\nr{non-relativistic }
\def\Nr{Non-relativistic }
\newcommand{\slPi}{/ {\hskip-0.27cm{\Pi}}}

\newcommand{\hh}{\rm h{\hskip-0.28cm{h}}}
\newcommand{\PP}{\rm P{\hskip-0.35cm{P}}}
\newcommand{\el}{\rm 1{\hskip-0.04cm{1}}}
\def\Lx{\varphi}
\newcommand{\tT}{\tilde{T}}
\newcommand{\tgamma}{\tilde{\gamma}}
\newcommand{\IIA}{\rm {I}\hskip-0.12mm{I}\hskip-0.2mm{A }}
\def\rmd{{\rm d}}
\def\lag{Lagrangian }\def\lags{Lagrangians }
\def\vs{\vskip 4mm}
\begin{document}
\begin{titlepage}
\begin{flushright}
UB-ECM-PF-05/34\\
Toho-CP-0578\\
hep-th/0512146
\end{flushright}
\vspace{.5cm}
\begin{center}
\baselineskip=16pt
{\LARGE Brane Dualities in \Nr Limit
}\\
\vfill
{\large  Kiyoshi Kamimura$^1$ and Toni Ramirez$^2$
  } \\
\vfill
{\small $^1$ Department of Physics, Toho University \\
Funabashi 274-8510, Japan 
\\ \vspace{6pt}
$^2$ Departament ECM, Facultat F{\'\i}sica,
  Universitat de Barcelona \\
  and\\
  CER for Astrophysics, Particle Physics and Cosmology, ICE/CSIC
\\
Diagonal 647, E-08028
      \\[2mm] }
\end{center}
\vfill
\begin{center}
{\bf Abstract}
\end{center}
{\small
\indent

We analyze brane dualities in the non-relativistic 
limit of the worldvolume actions. 
In particular we have analyzed how the \nr M$2$-brane is related via these 
dualities to \nr D$2$-brane, \nr IIA fundamental string and also, 
by using T-duality, to \nr D$1$-string.
These actions coincide with ones obtained from 
relativistic actions by taking non-relativistic limit, showing that the 
non-relativistic limit and the dualities commute in these cases.
 }\vspace{2mm} \vfill \hrule width 3.cm
\vspace{2mm}
{\footnotesize \noindent 
e-mails: kamimura@ph.sci.toho-u.ac.jp, tonir@ecm.ub.es}
\end{titlepage}
\addtocounter{page}{1}
\tableofcontents{}

\section{Introduction}

Non-relativistic limit of string theories appear to be solvable sectors 
\cite{Gomis:2000bd,Danielsson:2000gi}. 
It can provide us a deeper understanding of string
theories and can be helpful to answer some of open questions as 
for example AdS/CFT correspondence. 
In \cite{Brugues:2004an,Gomis:2004pw,Gomis:2005pg,Gomis:2005bj} the 
non-relativistic limits of strings and branes have been analyzed using
the  worldvolume actions. They have the Galilean supersymmetries and the 
\nr spectra. 
A notable property is that the kappa symmetry can be maintained in the 
\nr limit. By fixing the kappa gauge choice suitably these systems 
become free in the static gauge . 
    
In this paper we examine some of dualities between branes and 
strings in the non-relativistic limit using their worldvolume actions. 
It is interesting to see how these dualities remain in the non-relativistic 
limit\cite{Gomis:2000bd}. 
Here we study the dualities at the worldvolume actions
by considering the \nr M$2$-brane
as a starting point to obtain non-relativistic type IIA superstring and 
\nr D$2$-brane, using analogous procedures to the relativistic case
\cite{Duff:1987bx,Townsend:1995af}. 
We also examine the T-duality  between D$2$ and D$1$ branes 
\cite{Bergshoeff:1996cy,Green:1996bh} and find
that the T-duality transformations are compatible with 
the non-relativistic limit. 
In the cases we examine in this paper we see
the dualities remain 
in the non-relativistic sector of the string theories, ({\it Figure 1}).
\begin{figure}
\begin{center}
\includegraphics[scale=0.8]{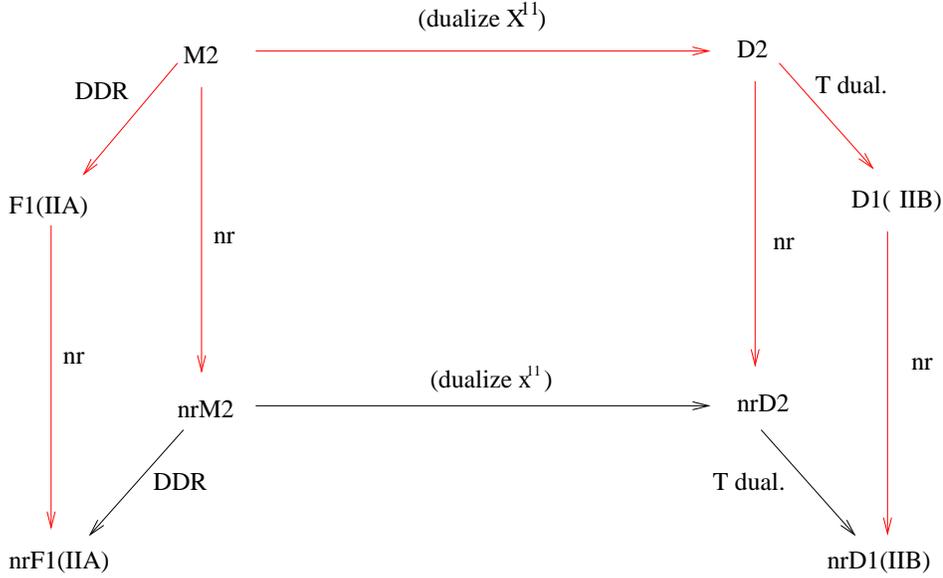}
\end{center}
\caption{\it This diagram shows the different ways in which the non-relativistic type IIA string, non-relativistic D$2$-brane and non-relativistic D$1$-string are obtained. Due to the commutativity of the non-relativistic limit and the dualities, we can arrive at the same actions by taking in different paths. }
\label{figure}
\end{figure}

The paper is organized as follows. 
In section $2$, we obtain the non-relativistic p-brane worldvolume action 
of Polyakov type. 
In section $3$ 
we consider a reduction from \nr M$2$-brane to \nr D$2$-brane by 
dualizing the $\el$-th coordinate in the same way as in the relativistic case. 
In section $4$ we discuss the double dimensional reduction of 
the non-relativistic M2 brane to obtain the non-relativistic type IIA 
superstring. In section $5$ we analyze the T-duality transformation of the 
non-relativistic D$2$-brane to non-relativistic D$1$-string. 
Summary is in the last section and
appendices are attached for notations and some detail.

\section{NR limit of p-branes}

The \nr limit of the p-branes have been discussed using their 
worldvolume actions for the bosonic branes\cite{Brugues:2004an} and the 
supersymmetric ones\cite{Gomis:2004pw}. It is established 
how to obtain worldvolume actions of \nr branes 
from the relativistic ones of Nambu-Goto form. 
Here we make a short summary but starting from Polyakov form action
and we find the \nr p-brane action of Polyakov form.

We start by considering the supersymmetric p-brane action in a flat
$D$ dimensional background,
\be
S^{p}=T_p\int d^{p+1}\xi\; {\cal L},\qquad 
{\cal L}=\CL^{P}\;+\;\CL^{WZ}\;+\;\CL^{B},
\label{LMp}
\ee 
where 
$T_p$ is the $p$-brane tension. The kinetic Lagrangian has the
Polyakov form using the worldvolume metric $\gamma_{IJ}, (I=0,1,...,p)$,
\bea
\CL^{P}&=&-\frac{\sqrt{-\gamma}}{2}\left(\gamma^{IJ}G_{IJ}-(p-1)\right).
\label{Lpt}
\eea
The WZ Lagrangian is the pullback of $p+1$ form $b$  
determined from a closed and superinvariant $p+2$ form $h$. For 
minimal spinor $\T$ in the $D$ dimensions, 
\bea
\CL^{WZ}\=b^*,\qquad h&=&db\;=\;-\frac{i}{p!}\;d\Tb\;\slPi^p\;d\T,
\label{hpp2f}\eea
where the super invariant one forms $\Pi^M, (M=0,...,D-1)$, 
and the induced metric $G_{IJ}$ are
\be
\Pi^M\= d X^M+{i}\Tb\Gamma^Md\T\equiv d\xi^I{\Pi_I}^M,
\qquad
G_{IJ}\={\Pi_I}^M{\Pi_{J}}^N\h_{MN}.
\ee
To show its closure we need the gamma matrix identity \bref{GammaI11} 
valid in the  $D$ dimensions where $p$-branes exist\cite{Bergshoeff:1987cm} 
(See Appendix \ref{app} for detailed notations). 
For the M$2$-brane we will discuss in this paper the explicit form of $b$ is
\bea
b&=&-
\frac{1}{2}(i\Tb\Gamma_{ML}d\T)\left( 
\Pi^M\Pi^L-
({i}\Tb\Gamma^{M}d\T)\Pi^L 
+\frac13({i}\Tb\Gamma^{M}d\T)({i}\Tb\Gamma^{L}d\T)\right). 
\label{bM2}\eea
$\CL^{B}$ is the $B$-field Lagrangian and is the pullback of 
a closed $B$-field. 
It is introduced without modifying the supergravity equations of motion. 
\vs

In ref.\cite{Gomis:2004pw} we have shown the \nr p-brane worldvolume \lag is
obtained from the relativistic one by rescaling of the supercoordinates 
and the tension as
\bea\label{NRrescaling}
X^\mu&=&\w x^\mu,\qquad X^A= x^A,\qquad 
\T=\sqrt{\w}\T_-+\frac{1}{\sqrt{\w}}\T_+,\qquad T_p= \w^{1-p}\tilde{T}_p.
\label{limit}\eea 
and then taking $\w \rightarrow \infty$. 
In the non-relativistic limit, the longitudinal coordinates  $X^\mu, 
(\mu=0,1,...,p)$  of the string are rescaled  while the transverse ones 
$X^A, (A=p+1,...,D-1)$ are left untouched, which implies that the transverse 
fluctuations are small.
 The scaling behavior  of the fermionic coordinates in \bref{limit} is 
characterized by their transformation properties under  the matrix  
$\Gamma_{*}$, which splits the fermionic coordinates into 
two eigenspaces $\T_\pm$ of $\Gamma_{*}$ with eigenvalues $\pm 1$,
\be
\T_\pm\equiv \PP_\pm\T,\qquad \PP_\pm=\frac12(1\pm\Gamma_*),\qquad
\Gamma_*=\Gamma_0...\Gamma_{p}.
\label{PProj}
\ee
In the Polyakov form we should also rescale the worldvolume metric 
$\gamma_{IJ}$ as
\bea
\gamma_{IJ}&=&\w^2\tgamma_{IJ}. 
\label{gamresc}\eea
In the $\w\to\infty$ limit the Lagrangian has divergent terms of order 
$\w^2$. 
They can be cancelled by adjusting the divergent 
contribution of the $B$-field Lagrangian $\CL^B$ in \bref{LMp}
in the \nr limit, 
\be
\CL^{B}\equiv{{T}_p}\det(\pa_IX^\mu)
=\w^2\tilde{T}_p\det(\pa_Ix^\mu)\equiv \w^2\;\tilde{T}_p\;\CL^{B}_{\rm div}.
\label{BLagM2}\ee
Its contribution to the energy is precisely 
compensated with the contribution coming from the tension of the brane.
Under these choices the action in the $\w \rightarrow \infty$ is finite 
and shown to have a \nr supersymmetry and kappa invariance. 
\vs

Making these rescalings in the \lag \bref{LMp} we get
\be
S^{p}=\tT_p\int d^{p+1}\xi\;\left(\w^2(\CL^{P}_{\rm div}+\CL^{WZ}_{\rm div}+
\CL^{B})+(\CL^{P}_{\rm fin}+\CL^{WZ}_{\rm fin})\right), 
\label{LMnr}
\ee 
where 
\bea
\CL^{P}_{\rm div}&=&- \frac{\sqrt{-\tgamma}}{2}\left(\tgamma^{IJ}g_{IJ}-(p-1)
\right),
\\
\CL^{P}_{\rm fin}&=&- \frac{\sqrt{-\tgamma}}{2}\;\tgamma^{IJ}G^{nr}_{IJ}
\eea
and \footnote{The symmetrization conventions here are 
${A_{(i}}B{_{j)}}=\frac12(A_iB_j+A_jB_i),\;
{A_{[i}}B{_{j]}}=\frac12(A_iB_j-A_jB_i)$.}
\bea\label{somedef}
g_{IJ}&=&{e_I}^\mu{e_J}^\nu\h_{\mu\nu},\qquad 
{e}^\mu=dx^\mu+{i}\Tb_-\Gamma^\mu d\T_-,
\nn\\
G^{nr}_{IJ}&=&{u_I}^A{u_J}^B\D_{AB}
+{2i}\Tb_+\Gamma_\mu{e_{(I}}^\mu\pa{_{J)}}\T_+,\quad 
{u}^A=dx^A+({i}\Tb_-\Gamma^Ad\T_++{i}\Tb_+\Gamma^Ad\T_-).
\nn\\
\eea
Here and hereafter inverse power terms of $\w$, which do not contribute
in the $\w\to\infty$ limit eventually, are omitted.  
The sum of divergent terms of the WZ  and $B$ field \lags is shown to be
\bea
\CL^{WZ}_{\rm div}+\CL^{B}&=&\det({e_I}^\mu)\=\sqrt{-g}.
\eea
By adding $\CL^{P}_{\rm div}$ the $\w^2$ terms of the total \lag is
\bea
\w^2\CL_{\rm div}&=&\w^2\left[
- \frac{\sqrt{-\tgamma}}{2}\left(\tgamma^{IJ}g_{IJ}-(p-1)\right)+\sqrt{-g}\;
\right].
\eea
It is  straightforward  to show that it has an expansion of 
$(\tgamma_{IJ}-g_{IJ})$ starting from their bi-linear terms,
\be
\w^2\CL_{div}=-\frac{\w^2}{2}\:A^{IJ,KL}(\tgamma_{IJ}-g_{IJ})
(\tgamma_{KL}-g_{KL}),
\label{supdivtot}\ee
where\footnote{We only need to consider $A$ and $A^{-1}$ in the vicinity of
${\tgamma_{IJ}=g_{IJ}}$. 
Corrections are proportional to $(\tgamma-g)^2$ and turn to vanish in the 
$\w\to\infty$ limit.}
\be
A^{IJ,KL}|_{\tgamma_{IJ}=g_{IJ}}\=\tT_p\frac{\sqrt{-g}}{4}(g^{IK}g^{JL}+
g^{IL}g^{JK}-g^{IJ}g^{KL}).
\label{supdivtot2}
\ee
We can now rewrite this superficially divergent term in a way that the $\w 
\rightarrow \infty$ limit can be 
taken smoothly. The idea is 
to introduce Lagrange  multipliers to rewrite  the 
action. For the case $p\neq 1$ it is
\be
\w^2\CL_{div}=\lambda^{IJ}\;(\tgamma_{IJ}-g_{IJ})
+{1\over 2\w ^2}A^{-1}_{IJ,KL}\lambda^{IJ}\lambda^{KL},
\label{lagrangep}
\ee
where 
\be
A^{-1}_{IJ,RS}A^{RS,KL}={\D_I}^{(K}{\D_J}^{L)},\quad 
A^{-1}_{IJ,KL}|_{\gamma_{IJ}=g_{IJ}}=\tT_p\frac{2}{\sqrt{-g}}
(g_{IK}g_{JL}+g_{IL}g_{JK}-\frac{2}{p-1}g_{IJ}g_{KL}).
\ee
\bref{lagrangep} reproduces \bref{supdivtot} by integrating out the 
$\lambda^{IJ}$ variables.
We can
 take  the strict non-relativistic limit  $\w \rightarrow
\infty$  in \bref{lagrangep} and be left with a finite contribution:
\be
\CL^*=\lambda^{IJ}(\tgamma_{IJ}-g_{IJ}).
\ee

For the $p=1$ case $A^{IJ,KL}$ is singular and a device
is required. Following to \cite{Gomis:2005pg} we write 
\be
\w^2\CL_{div}=-\w^2\:{\tT_p\over 2}\sqrt{-\tgamma}\tgamma^{00}
\tgamma_{\hmu\hnu}\;f^\hmu f^\hnu,
\label{supdivtotp1}\ee
where $\hmu=0,1$ and
\be
f^\hmu\equiv \left[ {e_0}^{\hmu}-{\sqrt{-\tgamma}\over \tgamma_{\el}}
\varepsilon^{\hmu\8\rho}\eta_{\8\rho\8\sigma}{e_1}^{\8\sigma}-{\tgamma_{01}
\over \tgamma_{\el}}{e_1}^{\hmu}\right]
\label{supdivtotp2}
\ee
and rewrite the superficially divergent term as 
\be
\CL_{div}=\lambda_{\hmu}f^\hmu
+{1\over 2\w ^2\tT_p\sqrt{-\tgamma}\tgamma^{00}}\lambda_\hmu\lambda_\hnu
\h^{\hmu\hnu},
\label{lagrange1}
\ee
which reproduces \bref{supdivtotp1} by integrating out the $\lambda_\hmu$ 
variables.
It becomes in the non-relativistic limit  $\w \rightarrow \infty$  
\be
\CL^*=\lambda_{\hmu}f^\hmu.
\ee
Now that we have properly defined the superficial divergence, 
we can finally write the complete finite 
action for non-relativistic p-branes as
\be
S^p_{nr}=\tT_p\int d^{p+1}\xi\;\left(
\CL^{P}_{\rm fin}+\CL^{WZ}_{\rm fin}+\CL^{*}\right)=\tT_p\int d^{p+1}\xi\;\left(-\frac{\sqrt{-\tgamma}}2\tgamma^{IJ}G^{nr}_{IJ}+\CL^{WZ}_{\rm fin}+\CL^{*}
\right) .\qquad 
\label{LMnrf}
\ee 
In contrast to the relativistic case, the \nr Polyakov Lagrangian \bref{LMnrf}
contains extra multiplier variables $\lam$'s. As was discussed in 
\cite{Gomis:2000bd} for the string case, they are indispensable to realize 
the \nr symmetry and the spectrum upon quantization.
\vs

The Nambu-Goto form of the \nr p-brane action can be obtained by
eliminating the worldvolume metric $\tgamma_{IJ}$ using the equations of motion
with respect to $\lam^{IJ}, (p\neq 1)$ or $\lam_\hmu, (p=1)$ cases,
\be
\tgamma_{IJ}=g_{IJ},\; (p\neq 1),\quad {\rm and} \quad  
\tgamma_{IJ}\propto g_{IJ},\; (p=1).
\ee
The action \bref{LMnrf} becomes
\bea
S^{p}_{nr}&=&\tT_p\int d^{p+1}\xi\;\left(
\CL^{NG}_{\rm fin}+\CL^{WZ}_{\rm fin}\right)\=
\tT_p\int d^{p+1}\xi\;\left(
- \frac{\sqrt{-g}}{2}\;g^{IJ}G^{nr}_{IJ}
+\CL^{WZ}_{\rm fin}\right).
\label{nrMpaction}
\eea
The finite part of the WZ Lagrangian $\CL^{WZ}_{\rm fin}$ is the pullback of
the $\w^{p-1}$-th order term $b_{nr}$ of  $p+1$ form $b$ in the \nr expansion. 
It is obtained by integrating  $\w^{p-1}$-th order term of $h$ in 
\bref{hpp2f}. The explicit form of $b_{nr}$ for the \nr M$2$-brane $(p=2)$ is 
given as \cite{Gomis:2004pw},
\bea
b_{nr}^{M2}&=& -{1\over2}\Bigg\{K^+_{\mu\nu}\left[e^\mu e^\nu - 
K_-^\mu e^\nu + {1\over3}K_-^\mu
K_-^\nu\right]
+  K^-_{\mu\nu}K_+^\mu \left[e^\nu - {1\over3}K_-^\nu\right]\nn\\
&& + \ L_{\mu B}\left[2e^\mu u^B - e^\mu L^B - K_-^\mu u^B + 
{2\over3}K_-^\mu L^B\right] \nn\\
&&+ \ K^-_{AB}\left[u^A u^B - L^A u^B + {1\over3}L^A L^B\right]\Bigg\},
\label{bM2nr}
\eea
where
\bea\label{KLdef}
K^\pm_{\mu\nu}& =& i \bar\T_\pm \Gamma_{\mu\nu} d\T_\pm, \qquad
K^\mu_\pm = i \bar\T_\pm \Gamma^\mu d\T_\pm, \qquad
K^-_{AB} = i \bar\T_- \Gamma_{AB} d\T_- , \nn\\
L_{\mu A} &=& i \bar\T_+ \Gamma_\mu \Gamma_A d\T_- +
i \bar\T_- \Gamma_\mu \Gamma_A d\T_+ , \qquad
L^A = i \bar\T_+ \Gamma^A d\T_- +i \bar\T_- \Gamma^A d\T_+. 
\eea

\vs
\section{\Nr M$2$ to \Nr D$2$ brane}

In \cite{Townsend:1995af} the D$2$-brane action is obtained by dualizing 
\el-th coordinate $X^{\el}$ of the M$2$-brane to the BI U(1) potential. 
In this section we show that the 
\nr D$2$-brane action is  obtained from \nr M$2$-brane 
by dualizing \el-th coordinate of the \nr M$2$-brane in the same manner
as in the relativistic case. 
As the \nr D$2$-brane action is obtained by taking the \nr limit of the 
relativistic one\cite{Gomis:2005bj}, the \nr limit and the dualization are
commutative. 
\vs


We first summarize briefly the result of 
\cite{Townsend:1995af}. 
In order to obtain  D$2$-brane from M$2$-brane, the U(1) gauge field degrees of 
freedom must be introduced. There is an extra coordinate $X^{\el}$ in the 
M2 brane which appears in the action in a form $\Lx=dX^{\el}$. 
By introducing one form Lagrange multiplier $A$ and adding a term 
${T_2}\Lx dA$ on the Lagrangian the $dX^{\el}$ can be replaced by $\Lx$.
The one form 
$\Lx$ can be regarded as an independent variable. 
The variation of the Lagrangian with respect to the $A$ gives an equation 
of motion $d\Lx=0$ whose solution is $\Lx=d\chi$. The M$2$-brane system
is reproduced by identifying the $\chi$ with $X^{\el}$.
On the other hand if we eliminate  the $\Lx$ by its equation of motion
we get a Lagrangian depending on the $A$. Since  the $A$ enters 
in the Lagrangian only through $F=dA$ the $A$ becomes a U(1) gauge 
field. The resulting Lagrangian is shown to describe
the D$2$-brane system.

More explicitly we start with the M$2$-brane action of Polyakov form
\bref{LMp}.
Replacing $\pa_I X^{\el}$ by $\Lx_I$, 
\bea
S^{M2}&=&-{T_2}\int d^3\xi\frac{ \sqrt{-\gamma}}{2}
\left\{\gamma^{IJ}\Pi^m_I\Pi^n_J\eta_{mn}+\gamma^{IJ}
\left(\Lx_I+i\bt\Gamma_{\el}\pa_I\T\right)
\left(\Lx_J+i\bt\Gamma_{\el}\pa_J\T\right)-1\right\}\nn\\
&&\hskip 10mm  +{T_2}\int \left\{-C^{(3)}-b^{(2)}\Lx\+\Lx\;d A \right\},
\eea
where the last term is the Lagrangian multiplier term and  $m=0,\dots,9$. 
The WZ three form $b$ of the M$2$-brane  \bref{bM2}
is split into a term $C^{(3)}$ independent of $dX^{\el}$ and a term 
linear in $dX^{\el}$ as 
\be
b\equiv -C^{(3)}-b^{(2)}dX^{\el},
\ee 
where
\bea
C^{(3)}&=&
\frac{1}{2}(i\Tb\Gamma_{mn}d\T)\left( 
\Pi^m\Pi^n-
({i}\Tb\Gamma^{m}d\T)\Pi^n 
+\frac13({i}\Tb\Gamma^{m}d\T)({i}\Tb\Gamma^{n}d\T)\right)
\nn\\
&+&
\frac{1}{2}(i\Tb\Gamma_{m}\Gamma_{\el}d\T)(\Pi^m-\frac{1}3i\Tb\Gamma^{m}d\T)
({i}\Tb\Gamma^{\el}d\T),
\\ 
b^{(2)}&=&
(i\Tb\Gamma_{m}\Gamma_{\el}d\T)(\Pi^m-\frac{1}2i\Tb\Gamma^{m}d\T).
\label{C3D2}
\eea
By integrating out the $\Lx_I$, the action becomes that of the D$2$-brane, 
 \bea
S^{D2}&=&-{T_2}\int d^3\xi\frac{ \sqrt{-\gamma}}{2}
\left\{\gamma^{IJ}\Pi^m_I\Pi^n_J\eta_{mn}-1
+\frac12\gamma^{IK}\gamma^{JL}\CF_{IJ}\CF_{KL}
\right\}\nn\\
&&\hskip 10mm  +{T_2}\int \left\{-C^{(3)}-
C^{(1)}\;\CF \right\},
\label{D2Poly}\eea
where 
\be
\CF=dA-b^{(2)},\qquad\textrm{and}\qquad C^{(1)}=i\bt\Gamma_{\el}d\T.
\ee
$\CF$ is the superinvariant U(1) field strength modified by the
NS two form $b^{(2)}$. $C^{(3)}$ and $C^{(1)}$ are the RR potential forms
in the flat supergravity background.
The second line of \bref{D2Poly} is giving the WZ Lagrangian of the D$2$-brane 
satisfying
\bea
d\left(C^{(3)}+C^{(1)}\CF\right)&=&
id\Tb\left(\frac{\Gamma_{mn}\Pi^m\Pi^n}{2!}+
\Gamma_{\el}\CF\right)d\T,
\eea
which is the condition required for the D$p$-brane actions 
\cite{Cederwall:1996pv,Aganagic:1996pe,Cederwall:1996ri,Bergshoeff:1996tu}.
\vs
 
\bref{D2Poly} is the D$2$-brane action of the Polyakov type. 
It has been shown\cite{Townsend:1995af}  perturbatively in $\CF$ 
that it gives the Dirac-Born-Infeld action by an integration of 
the worldvolume metric 
$\gamma_{IJ}$, 
\bea\label{D2DBI}
S^{BI}&=&-{T_2}\int d^3\xi\sqrt{-\det (G^{D2}_{IJ}+\CF_{IJ})}
-{T_2}\int\left(C^{(3)}+C^{(1)}\CF \right),
\eea
{\rm where $G^{D2}_{IJ}$ is the induced metric of D$2$-brane}
\be G^{D2}_{IJ}=\Pi^m_I\Pi^n_J\eta_{mn}.
\ee
In the Appendix \ref{apppoltong} we show it to all order of $\CF$.

\vs

The \nr D$2$-brane worldvolume action is obtained from that of
the \nr M$2$-brane.   
We begin with the non-relativistic action of M$2$-brane \bref{nrMpaction}.
In contrast to the relativistic case we can use the NG form in which
the worldvolume metric has been eliminated,
\be
S_{nr}^{M2}=-\tT_2\int d^3\xi \frac{\sqrt{-{g}}}{2}
\left\{g^{IJ}u^A_I
u^B_J\eta_{AB}+2i\bt_+\Gamma^{\mu}{e_\mu}^{I}\pa_{I}\T_+\right\}+
\tT_2\int b_{nr}^{M2}.
\ee
In the WZ Lagrangian the $b_{nr}^{M2}$ is given in \bref{bM2nr}  as 
\be
b^{M2}_{nr}=-C^{(3)}_{nr}-b^{(2)}_{nr}dx^{\el},
\ee
where $C^{(3)}_{nr}$ is the RR three form potential
and $b^{(2)}_{nr}$ is the NS two form in the \nr limit
\bea
C^{(3)}_{nr}&=&\frac12\Bigg\{K_{\mu\nu}^{+}\left[e^{\mu}e^{\nu}-
K_-^{\mu}e^{\nu}+\frac13K_-^{\mu}K_-^{\nu}\right]+K_{\mu\nu}^{-}K_+^{\mu}
\left[e^{\nu}-\frac13K_-^{\nu}\right]\nn\\
&&+L_{\mu b}\left[2e^{\mu}u^b-e^{\mu}L^b-K_-^{\mu}u^b+\frac23K_-^{\mu}
L^b\right]+K^-_{ab}\left[u^au^b-L^au^b+\frac13L^aL^b\right]
\nn\\
&&+L_{\mu \el}\left[e^{\mu}-\frac13K_-^{\mu}\right]L^{\el}
+K^-_{b\el}\left[u^b-\frac13 L^b\right]L^{\el}\Bigg\},
\label{defC3nr}
\nn\\
b^{(2)}_{nr}&=&L_{\mu \el}(e^\mu-\frac12K_-^\mu)+
K^-_{a \el}(u^a-\frac12L^a),
\label{defb2nr}\eea
where the fermionic bi-linears $K$'s and $L$'s are given in \bref{KLdef}.
We separate terms depending on $dx^{\el}$ and replace  $dx^{\el}$ with $\Lx$ 
by adding a Lagrange multiplier term $\Lx d\t A$ in the action
\bea
S_{nr}^{M2}&=&-\tT_2\int d^3\xi\frac{\sqrt{-{g}}}{2}
\left\{g^{IJ}{u_I}^a{u_J}^b\eta_{ab}+g^{IJ}(\Lx+C^{(1)}_{nr})_I
(\Lx+C^{(1)}_{nr})_J
+2i\Tb_+ \Gamma^{\mu}{e_\mu}^{I}\pa_{I}\T_+\right\}
\nn\\
&&+\tT_2\int \left(-C^{(3)}_{nr}-b^{(2)}_{nr}\Lx+\Lx\;d\t A\;\right),
\label{M2nr22}\eea
where $a,b=3,...,9$. $C^{(1)}_{nr}$ is RR one form potential equal to 
$L^{\el}$ in \bref{KLdef},
\be
C^{(1)}_{nr}=L^{\el}=  i \bar\T_+ \Gamma^{\el} d\T_- +i 
\bar\T_- \Gamma^{\el}d\T_+ 
\label{defC1nr}
\ee
and $(\Lx+C^{(1)}_{nr})_I$ is the $I$-th component of the one form
 $(\Lx+C^{(1)}_{nr})$ .
By integrating out the $\Lx_I$ we arrive at the \nr D$2$-brane action, 
 \bea
S_{nr}^{D2}&=&-{\7T_2}\int d^3\xi\frac{ \sqrt{-g}}{2}
\left\{g^{IJ}G^{D2}_{nr;IJ}
+\frac12g^{IK}g^{JL}\CF_{nr;IJ}\CF_{nr;KL}
\right\}
\nn\\&&\hskip 10mm  
+{\7T_2}\int \left\{-C^{(3)}_{nr}-C^{(1)}_{nr}
\;\CF_{nr} \right\},
\label{D2Poly2}\eea
where 
\be
G^{D2}_{nr;IJ}={u_I}^a{u_J}^b\eta_{ab}+2i\bt_+\Gamma_{\mu}
{e_{(I}}^{\mu}\pa_{J)}\T_+. 
\label{defGnrD2}\ee
$\CF_{nr}$ is the superinvariant U(1) field strength modified by the
NS two form $b^{(2)}_{nr}$,
\be
\CF_{nr}=d\7A-b^{(2)}_{nr}.
\label{defCFnr}\ee
\vs

The \nr D$2$-brane action \bref{D2Poly2} obtained by dualizing $x^{\el}$ of
 the \nr M$2$-brane
coincides with the one obtained by taking the \nr limit of the 
D$2$-brane action.
The \nr D$p$-branes have been derived \cite{Gomis:2005bj} 
 by taking the \nr limit of the DBI actions for general $p$. 
Here we illustrate it for
the D$2$-brane case but starting from the Polyakov form action \bref{D2Poly}.
To take the \nr limit of the D$2$-brane action \bref{D2Poly} we make the 
rescaling  as in the M$2$-brane case, \bref{NRrescaling} and \bref{gamresc}
\be
X^{\mu}=\w x^{\mu},~~~~~~~X^a=x^a, ~~~~~~~ 
\T=\sqrt{\w}\T_-+\frac1{\sqrt\w}\T_+,
\nn\ee
\be
\gamma_{IJ}=\w^2\tgamma_{IJ},~~~~~~~~~~~T_2=\frac1{\w}\tilde{T_2},
\label{gamresc2}\ee
where $\T_\pm$ is defined using the projectors for D$2$-brane,
\be
\PP_\pm=\frac12(1\pm\Gamma_*),\qquad
\Gamma_*=\Gamma_0\Gamma_{1}\Gamma_{2}.
\label{PProjD2}
\ee
In addition to it
we need the rescaling of the U(1) gauge field, 
\be
A_I=\w \7A_I.
\label{Aresc2}\ee
These rescalings in \bref{D2Poly} lead us to
\be
S^{D2}=\tT_2\int d^3\xi\left[\w^2 \left(\cL_{\textrm{div}}^P+
\cL_{\textrm{div}}^{WZ}\right)+\left(\cL_{\textrm{fin}}^P+\cL_{\textrm{fin}}^{WZ}\right)\right],
\label{LagD2Poly}\ee
where
\bea \label{pdivL}
\cL_{\textrm{div}}^P=-\frac{\sqrt{-\tgamma}}2\left(\tgamma^{IJ}g_{IJ}-1\right),
\quad
\cL_{\textrm{fin}}^P=-{\sqrt{-\tgamma}}\left(
\frac12\tgamma^{IJ}G^{D2}_{nr;IJ}+\frac14\tgamma^{IK}\tgamma^{JL}
\CF_{nr;IJ}\CF_{nr;KL}
\right),
\nn\\
\eea
where $G^{D2}_{nr;IJ}$ and $\CF_{nr;IJ}$ ere  given in \bref{defGnrD2} 
and \bref{defCFnr}.
 The divergent terms have the same form as M$2$-brane case, 
except the dimensions of the target space is 10 in the D$2$-brane.
Here we make an alternative analysis to obtain the NG form of action.
In the above Lagrangian \bref{LagD2Poly} 
we regard $\w$ sufficiently large but finite 
and inverse power terms of $\w$, which will not contribute, are omitted. 
Now, we remove the dependence on the worldvolume metric  $\tgamma_{IJ}$ 
from the Lagrangian \bref{LagD2Poly}  
by solving the equation of motion of $\tgamma_{IJ}$ as
\be
\tgamma_{IJ}=g_{IJ}+\frac1{\w^2}T_{IJ},
\label{tgammaD2}
\ee
where $T_{IJ}$ is the worldvolume
energy-momentum tensor defined from $\cL_{\textrm{fin}}$.
The  $\w^{-2}$ term in $\tgamma_{IJ}$ could give finite contributions 
when the $\tgamma_{IJ}$ in \bref{tgammaD2}  
is introduced back into the action \bref{LagD2Poly}.
However, it can be shown that these terms cancel and 
do not contribute after all\footnote{It also shows that the 
sub-leading terms
in \bref{gamresc} and  \bref{gamresc2},  if any, do not contribute to
the finite Lagrangians. 
See Appendix \ref{appprof}.}. 
For further calculations we can take only the leading term of \bref{tgammaD2} ;
$\tgamma_{IJ}=g_{IJ}$.

Using it in the divergent term $\cL_{\textrm{div}}^P$ it becomes
\be
d^3\xi\cL_{\textrm{div}}^P=-d^3\xi\sqrt{-g}=\frac1{3!}\varepsilon_{\mu\nu\rho}e^\mu 
e^\nu e^\rho.
\ee
Combining it with the divergent term of the WZ Lagrangian
$\cL_{\textrm{div}}^{WZ}$ of \bref{D2Poly}
\bea
d^3\xi\cL_{\textrm{div}}&=&\frac1{3!}\varepsilon_{\mu\nu\rho}e^\mu e^\nu e^\rho
-\frac12\left\{K_{\mu\nu}^{-}\left[e^{\mu}e^{\nu}-K_-^{\mu}e^{\nu}
+\frac13K_-^{\mu}K_-^{\nu}\right]\right\}
\nn\\
&=&\frac1{3!}\varepsilon_{\mu\nu\rho}\;dx^\mu dx^\nu dx^\rho.
\eea
It is an exact form and is canceled by adding the $B$ field 
Lagrangian same form as \bref{BLagM2}.
Now, we can take the limit $\w\rightarrow\infty$ and obtain the 
non-relativistic action for D$2$-brane,
 \bea
S_{nr}^{D2}&=&-{\7T_2}\int d^3\xi\frac{ \sqrt{-g}}{2}
\left\{g^{IJ}G^{D2}_{nr;IJ}
+\frac12g^{IK}g^{JL}\CF_{nr;IJ}\CF_{nr;KL}
\right\}
\nn\\&&\hskip 10mm  
+{\7T_2}\int \left\{-C^{(3)}_{nr}-C^{(1)}_{nr}
\;\CF_{nr} \right\}.
\label{D2Poly3}\eea
It has the same form as \bref{D2Poly2} obtained by 
dualizing $x^{\el}$ of the \nr M$2$-brane. 

\vs
\section{\Nr M$2$-brane to \Nr IIA superstring}

The reduction of the M$2$-brane to the IIA superstring has been discussed 
\cite{Duff:1987bx}. We first make a short review of the relativistic case.
We assume the M$2$-brane extends in 
$X^0,X^1$ and $X^{\el}$ directions and are the longitudinal coordinates. 
In doing the reduction of the M$2$-brane to the
\IIA superstring 
we assume the $X^{\el}$ is a compact coordinate parametrized by $\rho$
\be
X^{\el}(\tau,\s ,\rho)=\rho
\ee
and other  supercoordinates are independent of it, $(m=0,1,...,9)$
\be
X^m(\tau,\s,\rho)=X^m(\tau,\s ),
\qquad \T(\tau,\s ,\rho)=\T(\tau,\s ).
\ee
Using this $(j=0,1)$
\be
\Pi^{\el}_j={i}\Tb\Gamma^{\el}\pa_j\T, 
\qquad \Pi^{\el}_2=\pa_2X^{\el}=1 
\ee
and the $3\times 3$ induced metric of the M$2$-brane becomes
\bea
G_{IJ}=
\pmatrix{ {\Pi_i}^{m}{\Pi_j}^n\h_{mn}+{\Pi_i}^{\el}\Pi_{\el,j}&{\Pi_i}^{\el}\cr
{\Pi_j}^{\el}&1}
\eea
It follows
\be
G=\det(G_{IJ})=\det(G^{\IIA}_{ij})=G^{\IIA},
\label{ngIIA}\ee
where $G^{\IIA}_{ij}$ is  $2\times 2$ induced metric of the \IIA string
\be
G^{\IIA}_{ij}={\Pi_i}^{m}{\Pi_j}^n\h_{mn}.
\ee
Using them the WZ term of the  M$2$-brane  \bref{bM2} becomes the \IIA 
Green-Schwarz superstring action\cite{Green:1983wt}
\bea
d\xi^3\CL^{M2;WZ}&=&d\rho\,d\xi^2\,
\varepsilon^{ij}(i\Tb\Gamma_{m}\Gamma_{\el}\pa_i\T)\left( 
\Pi^m_j-
\frac12({i}\Tb\Gamma^{m}\pa_j\T) \right)=d\xi^3\CL^{\IIA;WZ}.
\label{wzIIA}
\eea
Then the M$2$-brane action becomes, using \bref{ngIIA} and \bref{wzIIA},
the \IIA superstring action 
\bea
S^{\IIA}&=&T^{\IIA}\int d\tau d\s  
(-\sqrt{-G^{\IIA}} +\;\CL^{\IIA;WZ})
\eea
with the string tension
\be
T^{\IIA}=(\int d\rho)\;T^{M2}.
\ee

\vs

The \nr \IIA superstring action has been obtained from the Green-Schwartz
action in \cite{Gomis:2004pw}. 
Here we show it is obtained by the double dimensional reduction 
of the \nr M$2$-brane. The reduction is similar to the relativistic case
given above but an attention should be paid for the spinor sector.
The \el D fermion $\T$ is split into the Majorana-Weyl spinors $\T^\pm$  
of chiralities $\pm1$ using $\;\hh^\pm=\frac12(1\pm\Gamma_{\el})$. 
In the \nr limit each Majorana-Weyl spinors are further separated 
using the projection operators $\PP_\pm$ in 10D.
Since the longitudinal coordinates of the M$2$-brane are $x^0,x^1$ and 
$x^{\el}$ the projection operators
$\PP_\pm$ in \bref{PProj} of the \nr M$2$-brane is 
\be
\PP_\pm=\frac12(1\pm\Gamma_*),\qquad
\Gamma_*=\Gamma_0\Gamma_1\Gamma_{\el}.
\ee
A crucial property is that the $\Gamma_*$ has same form for the 
\IIA superstring and the projection operators, $\;\hh_\pm$ and $\;\PP_\pm$, 
are commuting.
It makes the \nr limit and the double dimensional reduction commutative.

In doing the reduction of the \nr M$2$-brane to the \nr
\IIA superstring we follow the same procedure as the relativistic case.
We assume $x^0,x^1$ and $x^{\el}$ are the longitudinal coordinates
and the $x^{\el}$ is a compact coordinate parametrized by $\rho$
\be
x^{\el}(\tau,\s ,\rho)=\rho.
\ee
Other supercoordinates are independent on it, 
\be
x^m(\tau,\s,\rho)=x^m(\tau,\s ),
\qquad 
\T_\pm(\tau,\s ,\rho)=\T_\pm(\tau,\s ).
\ee
Under these assumptions, for $(i=0,1;\;\hmu=0,1)$, 
\bea
\pmatrix{{e_{i}}^{\hmu},&{e_{2}}^{\hmu}\cr{e_{i}}^{\el},& {e_{2}}^{\el} } &=&
\pmatrix{\pa_ix^{\hmu} +{i}\Tb_-\Gamma^{\hmu}\pa_i\T_-, &0\cr{i}\Tb_-
\Gamma^{\el}\pa_i\T_-, &1 },
\nn\\
g_{IJ}&=&\pmatrix{g^{\IIA}_{ij}+
{e_{i}}^{\el}{e_{j}}^{\el},&{e_{i}}^{\el} \cr
{e_{j}}^{\el}& 1}, \qquad
g^{\IIA}_{ij}\equiv {e_{i}}^{\hmu}{e_{j}}^{\hnu}\h_{\hmu\hnu},
\\
g^{IJ}&=&\pmatrix{{g_{\IIA}^{ij}},&-{g_{\IIA}^{ik}}{e_k}^{\el} \cr
-{g_{\IIA}^{jk}}{e_k}^{\el}& 1+{g_{\IIA}^{k\l}}{e_k}^{\el}{e_\l}^{\el}}, 
\qquad
\det{g_{IJ}}=\det{g^{\IIA}_{ij}}.
\eea
where $g_{\IIA}^{ij}$ is the inverse of induced metric $g^{\IIA}_{ij}$
of the \IIA superstring
.
For the transverse components, with the index $\ha=2,...9,$ 
\bea
{u_{i}}^{\ha}=\pa_ix^{\ha}+({i}\Tb_-\Gamma^{\ha}\pa_i\T_++{i}\Tb_+\Gamma^{\ha}\pa_i
\T_-),\qquad
{u_{2}}^{\ha}=0,
\eea
\bea
G_{IJ}^{nr}=\pmatrix{
{u_i}^{\ha}{u_j}^{\hb}\D_{\ha\hb}+{2i}\Tb_+\Gamma_{\hmu}{e_{(i}}^{\hmu}\pa{_{j)}}\T_+
+{2i}\Tb_+\Gamma_{\el}{e_{(i}}^{\el}\pa{_{j)}}\T_+,&
{i}\Tb_+\Gamma_{\el}\pa{_{i}}\T_+ \cr
{i}\Tb_+\Gamma_{\el}\pa{_{j}}\T_+, & 0}.
\label{nrGij}
\eea
Using them in the Nambu-Goto term of the \nr M$2$-brane 
\lag \bref{nrMpaction} it becomes that of the \nr \IIA superstring,
\bea
\CL^{NG}_{\rm fin}&=&- \frac{\sqrt{-g}}{2}\;g^{IJ}G^{nr}_{IJ}=
- \frac{\sqrt{-g^{\IIA}}}{2}\;g^{ij}_{\IIA}G^{\IIA}_{nr;ij},
\nn\\&&
G^{\IIA}_{nr;ij}={u_i}^{\ha}{u_j}^{\hb}\D_{\ha\hb}
+{2i}\Tb_+\Gamma_{\hmu}{e_{(i}}^{\hmu}\pa{_{j)}}\T_+.
\label{nrM2action3}
\eea
In the WZ three form $b^{M2}_{nr}$ of the \nr M$2$-brane \bref{bM2nr} 
only terms including ${e_2}^{\el}=1$ remain and
\bea
b^{M2}_{nr} 
= -d\rho\Bigg\{
K^+_{\hmu {\el}}({e}^{\hmu} -\frac12K_{-}^{\hmu})
+  \frac12{K^-_{\hmu {\el}}}K_{+}^{\hmu} 
+L_{\hb\el}\left(u^{\hb}-\frac12 L^{\hb}\right) 
\Bigg\}.
\label{bM2nr2}
\eea
Using a relation 
$\Gamma_{\hmu}\Gamma_{\el}\T_{\pm}=\pm\varepsilon_{\hmu\hnu}\Gamma^{\hnu}
\T_{\pm}$ 
in \bref{gami2} and the \IIA cyclic identity \bref{cyclicIIA} 
it becomes the WZ \lag of the \nr \IIA 
superstring given in \cite{Gomis:2004pw} up to an exact form. 
Then the \nr M$2$-brane action is reduced to that of the \nr \IIA superstring 
\bea
S^{\IIA}_{nr}&=&\tT^{\IIA}\int d\tau d\s \left\{
 -{e\over2}\;  g^{jk}_{\IIA}\;{u_j}^{\ha} {u_k}^{\hb}\D_{\ha\hb}\;-\;2i\;
e\,(\Tb_+\Gamma^{\hmu}{e_{\hmu}}^i\pa_i\T_+) \right.
\nonumber\\
&&\left. - 2i\;\vep^{jk}
\left(\Tb_+\Gamma_{\ha}\Gamma_{\el}\;\pa_j\T_-\right)
\left(u_k{}^{\ha} -{i}\;\Tb_+\Gamma^{\ha}\;\pa_k\T_-\right)\right\},
\label{Lag22}
\eea
where $e$ and ${e_{\hmu}}^i$ are the determinant and inverse of 
${e_i}^{\hmu}$ respectively and 
\be
 \tT^{\IIA}=(\int d\rho) \;\tT^{M2}.
\ee
\vs

\section{T-duality of \nr D$2$ and D$1$ branes}

It is known D$p$-branes are T-dual of D$(p+1)$-branes 
\cite{Bergshoeff:1996cy,Green:1996bh}. The T-duality covariance of the 
worldvolume actions of D-branes is studied in 
detail\cite{Simon:2000bi,Kamimura:2000bi}. 
The T-dualities of the \nr branes have been discussed in \cite{Gomis:2000bd}.
In this section we discuss the T-duality as that of the 
\nr D-brane worldvolume actions. 
Although our results can be applied to general D$p$-branes we 
confine our discussions to the  case of the T-duality between D$2$ and D$1$
branes for definiteness. 
There is a subtlety in the \nr limit since the D$2$-brane belongs to 
\IIA  sector while D$1$-string does to IIB sector and there is a chirality 
flip of a Majorana-Weyl fermion. 
In contrast to the case of reduction from the M$2$ to the \IIA superstring
discussed in the last section, 
the chirality
projectors $\;\hh^\pm$ and \nr projectors $\;\PP_\pm$ do not commute. 
Nevertheless we can show that the T-duality transformations and \nr limits
are commutative and   
the \nr D$1$-string can be obtained from \nr D$2$-brane by the T-duality.  
 \vs

{\underline{D$2$ to D$1$}}
\vs

In the relativistic D$2$-brane $X^0,X^1$ and $X^{2}$ are the longitudinal 
coordinates and  depend on $\xi^I=(\tau,\s,\rho)$.
In doing reduction to D$1$ we assume one of the longitudinal coordinate, $X^2$, 
is compact and parametrized by $\rho$ 
\be
X^{2}(\tau,\s ,\rho)=\rho
\ee
while others $X^m,(m\neq 2), A_I$ and $\T$ are functions of 
$\tau$ and $\s$ only.  
The dynamical degree of freedom, ${X'}^2$, of D$1$ coordinate is supplied 
from that of a U(1) gauge field component $A_2$,
\footnote{
In this section primed variables are those of D$1$ and unprimed ones are 
of D$2$. The variables in the \nr limit are indicated with tildes.}
\be
{{X'}^2}\={A_2},
\ee 
and others are
\be 
 {X'}^m\=X^m,\quad (m\neq 2) \qquad 
A'_j\=A_j,\quad (j=0,1).
\ee
The D$2$(IIA) spinors are related to those of D$1$(IIB) by flipping one of 
chirality component
\bea
\T'&=&\pmatrix{{\T'}^1\cr{\T'}^2 }\=\left(
\frac{1+\tau_3}2+\frac{1-\tau_3}2\Gamma_2\right)
\pmatrix{ \T^+\cr 
\T^-}
\=\pmatrix{\;\hh^+ \T\cr 
\Gamma_2\;\hh^-\T}.
\eea
Both IIB spinors ${\T'}^1$ and ${\T'}^2$ are Majorana-Weyl fermions with same 
chirality $+$.
Under this T-duality transformation the worldvolume action of the D$2$-brane 
\bref{D2DBI} turns out to be that of the D$1$-string, 
\bea
\label{D1DBI}
S^{D1}&=&-{T_1}\int d^2\xi\sqrt{-\det (G^{D1}_{ij}+\CF'_{ij})}
-{T_1}\int\;C^{(2)},
\eea
where 
\be G^{D1}_{ij}={\Pi'}^m_i{\Pi'}^n_j\eta_{mn},\qquad 
{\Pi'}^m\=d{X'}^m+i\Tb'\Gamma^m\T',
\nn\ee
\be
\CF'=dA'-i\Tb'\Gamma^m\tau_3\T'(d{X'}^m+\frac{i}{2}\Tb'\Gamma^m\T'),
\quad
C^{(2)}\=i\Tb'\Gamma^m\tau_1\T'(d{X'}^m+\frac{i}{2}\Tb'\Gamma^m\T').
\ee
Although the transformation 
breaks manifest Lorentz covariance the resulting action is superPoincare
invariant and kappa invariant\cite{Simon:2000bi,Kamimura:2000bi}.
\vs

{\underline{  D$1$ to \nr D$1$ }}
\vs
In the \nr limit the D$1$-string variables are rescaled as 
\be
 {\widetilde{({x'}^\hmu)}}\=\frac{1}{ \w}{X'}^\hmu,  \quad (\hmu=0,1), \qquad 
{\widetilde{(A'_j)}}\=\frac{1}{ \w}A'_j,\quad (j=0,1),
\ee
\bea
{\widetilde{(\T')}}&=&\left(
\sqrt{\w}\;\PP\;'_++\frac{1}{\sqrt{\w}}\;\PP\;'_-\right)
\T',\qquad \7T_1\= \; T_1,
\label{tpTheta}\eea
where $\PP\;'_\pm$ are the \nr projectors for D$1$-string,
\be
\PP\;'_\pm=\frac12(1\pm{\Gamma'}_{*}),\qquad {\Gamma'}_{*}=\Gamma_{01}\tau_1.
\ee
By taking $\w\to\infty $ limit we obtain \nr D$1$-string 
action\cite{Gomis:2005bj}. Abbreviating tilde and prime indices for the
\nr D1 variables,  
\bea
S^{D1}_{nr}&=&\tT^{D1}\int d\tau d\s \left\{
 -{e\over2}\;  g^{jk}\;{u_j}^{\ha} {u_k}^{\hb}\D_{\ha\hb}\;-\;2i\;
e\,(\Tb_+\Gamma^{\hmu}{e_{\hmu}}^i\pa_i\T_+)-
\frac{e}{4}g^{ij}g^{k\l}\CF_{nr;ik}^{D1}\CF_{nr;j\l}^{D1} \right.
\nonumber\\
&&\left. - 2i\;\vep^{jk}
\left(\Tb_+\Gamma_{\ha}\tau_{1}\;\pa_j\T_-\right)
\left(u_k{}^{\ha} -{i}\;\Tb_+\Gamma^{\ha}\;\pa_k\T_-\right)\right\}.
\label{D1nrlim}
\eea
Comparing to the \nr \IIA superstring \bref{Lag22}
$\Gamma_{\el}$ is replaced by $\tau_1$ in the WZ term and 
$\CF_{nr}^{D1}$ is  $\CF_{nr}$ of \bref{defCFnr} in which $\Gamma_{\el}$
is replaced by $\tau_3$.

\vs
{\underline{  D$2$ to \nr D$2$}}
\vs
On the other hand starting from the relativistic D$2$-brane action
\bref{D2DBI} we have obtained the \nr D$2$-brane action
\bref{D2Poly3} in the \nr limit. 
The \nr rescalings are \bref{gamresc2} and \bref{Aresc2},
\be
 {\7x}^\mu\=\frac{1}{\w}\;X^\mu \;, \quad (\mu=0,1,2),\quad
\t A_I\= \frac{1}{\w}\; A_I, \quad (I=0,1,2)
\ee
and   
\be
\7\T=(\sqrt{\w}\;\PP_++\frac{1}{\sqrt{\w}}\;\PP_-)\T,
\qquad \7T_2\= \w\; T_2,
\ee
where $\PP_\pm$ are the projectors for D$2$-brane,
\be
\PP_\pm=\frac12(1\pm{\Gamma}_{*}),\qquad {\Gamma}_{*}=\Gamma_{012}.
\ee

\vs
{\underline{ \Nr D$2$ to \nr D$1$ }}
\vs
From the  \nr D$2$-brane to \nr D$1$-string we make the same T-duality 
transformation as the relativistic case. Compactify in 
the longitudinal direction $\7x^{2}$
and choose
\be
\7x^{2}(\tau,\s ,\rho)=\rho
\ee
while others are independent of  $\rho$. 
The dynamical degree of freedom, ${(\7x^2)}{'}$, of \nr 
D1 string coordinate is transferred 
from that of a U(1) gauge field component $\7A_2$,
\be
{(\7x^2)}{'}\={\7A_2},
\ee 
and others are
\be 
 {(\7x^m)'}\=\7 x^m,\quad (m\neq 2), \qquad 
 (\7A_j)'\=\7A_j,\quad (j=0,1).
\ee
The \nr D$2$(IIA) spinors are related to those of \nr 
D$1$(IIB) by flipping one of 
chirality component 
\bea
(\7\T)'&=&
\pmatrix{\;\hh^+ \7\T\cr 
\Gamma_2\;\hh^-\7\T}
\=
\pmatrix{\;\hh^+ (\sqrt{\w}\;\PP_++\frac{1}{\sqrt{\w}}\;\PP_-)\T\cr 
\Gamma_2\;\hh^-(\sqrt{\w}\;\PP_++\frac{1}{\sqrt{\w}}\;\PP_-)\T}.
\label{ptTheta}\eea
\vs

They show that the T-duality transformations and the \nr limit
are commutative especially 
\be
(\7\T)'\=\widetilde{(\T')}\qquad {\rm and}\qquad
{(\7 x^2)'}\=\widetilde{({x^2}')}.
\ee
Note the latter comes from the fact that both the longitudinal
coordinates $X^\mu$ and the U(1) gauge field $A_I$ are rescaled in
a same power of $\w$ under the \nr rescaling.
The commutativity guarantees that the \nr D$1$-string action obtained by 
the T-duality transformation of the \nr D$2$-brane
coincides with that given by the \nr limit of 
the D$1$-string. 
Actually making the transformations on the \nr D2 brane action
\bref{D2Poly3} we can obtain  \nr D1 string action
\bref{D1nrlim}.

\vs 

\section{Summary}

In this paper we have analyzed, at the world volume level, several
corners of M-theory described by non-relativistic theories. In particular
we have studied how these non-relativistic theories are mapped to each
other by the duality symmetries of the worldvolume actions. In particular
we have analyzed how the \nr M$2$-brane is related via these dualities to
\nr D$2$-brane, \nr IIA fundamental string and also by using T-duality to 
\nr D$1$-string. Their worldvolume actions coincide with those obtained from 
the relativistic actions by taking the non-relativistic limit 
\cite{Gomis:2004pw,Gomis:2005bj}. 
Thus we have shown that the non-relativistic limit and the 
duality transformations are commutative as illustrated in the {\it Figure 1}.

From a more technical point of view we have constructed \nr
branes starting from the Polyakov formulation of relativistic M$2$ and 
D$2$-brane. One can also show that the kappa symmetry is maintained in 
the non-relativistic limit and  it is also preserved
through the use of duality symmetries of M theory.
Since non-relativistic D-branes in the static gauge and $\theta_-=0$
are described by free supersymmetric gauge theories it would be
interesting to see how these M dualities are realized on these
supersymmetric gauge theories.

\vskip 1cm
\noindent
\section*{Acknowledgments.}
\noindent We are grateful to Joaquim Gomis and Jan Brugues for valuable
discussions. This work was partially funded by the European Community's Human
Potential Programme under contract MRTN-CT-2004-005104 `Constituents,
fundamental forces and symmetries of the universe',
the Spanish grant MYCY FPA 2004-04582-C02-01, and by the
Catalan grant CIRIT GC 2001, SGR-00065.

Toni Ramirez thanks the Cosmology and Particle physics group of
Toho University for the warm hospitality.

\vskip 1cm

\begin{appendix}
\section{Notations and some useful formulae}\label{app}

Here we summarize some notations. Indices are
\[
\begin{array}{ccc}~~~~~~
&{\underline{ 10~{\rm dim}}} & {\underline{ \el~{\rm dim}}}\nn\\
{\rm target\; space}:& m,n=0,...,9 & M,N=0,...,10\nn\\
\nn\\~~~~~~
&{\underline{ p=1}} & {\underline{ p=2}} \nn\\
{\rm target\; space, longitudinal}: &\hmu,\hnu=0,1 & \mu,\nu=0,1,2\nn\\
{\rm target\; space, transverse}:& \ha,\hb=2,...,9 & a,b=3,...,9\nn\\
{\rm worldvolume}:& i,j=0,1 & I,J=0,1,2
\end{array}
\]
and we use $A,B=3,\dots,9,\el$ for transverse indices 
when we work in $\el$ dimensions.
The metrics of target space and worldvolume have mostly
$+$ signatures. The totally antisymmetric Levi-Civita tensor is normalized by
$\varepsilon^{012...p}=+1$, and $\varepsilon _{012...p}=-1$.

The gamma matrices are $\Gamma^m$ and 
$\Gamma_{\el}=\Gamma_{0}\Gamma_{1}...\Gamma_{9}$.
They can be chosen real by taking the charge conjugation matrix 
$C=\Gamma_0$ and satisfy
\be
C^{-1}\Gamma^MC\=-(\Gamma^M)^T,\qquad C^T=-C.
\ee
The conjugate $\Tb$ of the Majorana spinor is defined by $\Tb=\T^TC$.  
For type IIA theories $\theta$ is a  Majorana spinor while for type IIB
theories there are two Majorana-Weyl spinors $\T_\alpha$ ($\alpha=1,2$)
of the same chirality. The index $\alpha$ on which 
the Pauli matrices $\tau_1, \tau_2,\tau_3$ act 
 is not displayed explicitly.
 This leads to some
useful symmetry relations as
\begin{equation}
\bar \chi \lambda =\bar \lambda \chi ,\qquad   \lambda =\Gamma _m\epsilon
\ \rightarrow \ \bar \lambda =-\bar \epsilon \Gamma
  _m,\qquad \lambda =\Gamma _{\el}\epsilon \ \rightarrow\ \bar \lambda =-\bar \epsilon \Gamma
  _{\el}.
 \label{barGamma}
\end{equation}

In $\el$-dimensions we have the gamma matrix identity
\bea
(C\Gamma_{ML})_{(\A\B}(C\Gamma^{L})_{\gamma\D)}&=&0,
 \label{GammaI11}\eea
where  all  indices $\A,\B,\gamma,\D$
are symmetrized. Therefore
\bea
(d\Tb\Gamma_{ML}d\T) (\Tb\Gamma^Ld\T)=
-(\Tb\Gamma_{ML}d\T) (d\Tb\Gamma^Ld\T).
\eea
In $10$-dimensions the gamma matrix identity is
\be
(C\Gamma_m\Gamma_{\el})_{\A(\B}(C\Gamma^{m})_{\gamma\D)}+
(C\Gamma_m)_{\A(\B}(C\Gamma^{m}\Gamma_{\el})_{\gamma\D)}=0,
\label{cyclicIIA}\ee
where $\Gamma_{\el}$ can be replaced by $\tau_i$ ($i=1,3$) for type 
IIB spinors.

From this gamma matrix identity we can derive cyclic identities 
in $10$-dimensions
\begin{eqnarray}
 \sum_{IJK\;\mathrm{cyclic}}\left[\Gamma_m
\T_I\;(\bt_J \Gamma^m\;\T_K)+ \Gamma_m \Gamma_{\el}\T_I\;(\bt_J
\Gamma^m\Gamma_{\el}\;\T_K)\right]=0, \label{cyclic2}
\end{eqnarray}
for the type IIA spinors and
\begin{equation} \sum_{IJK\;\mathrm{cyclic}} \left\{
\Gamma_m\tau_i\T_I\left(\bt_J \Gamma^m \T_K\right) +
\Gamma_m\T_I\left(\bt_J \Gamma^m \tau_i \T_K\right)\right\} = 0,\qquad
(i=1,3).
\label{CIDB}
 \end{equation}
for type IIB spinors.

In the \nr brane theories spinors are separated by 
projection operators 
\be 
\PP_{\pm}=\frac12\left(1\pm\Gamma_*\right),~~~~~~~~~~~~~~~~
\T_{\pm}=\PP_{\pm}\T
\ee
where $\Gamma_*$ is defined for each systems as
\bea
\rm{IIA~string:}&& \Gamma_*=\Gamma_0\Gamma_1\Gamma_{\el}\nn\\
\rm{IIB~string:}&& \Gamma_*=\Gamma_0\Gamma_1\tau_3\nn\\
\rm{D1:}&&\Gamma_*=\Gamma_0\Gamma_1\tau_1\nn\\
\rm{D2:}&&\Gamma_*=\Gamma_0\Gamma_1\Gamma_2\nn\\
\rm{M2:}&&\Gamma_*=\Gamma_0\Gamma_1\Gamma_{\el}.
\eea

For  the D$2$-brane there is a relation
\begin{equation}\label{D2}
\Gamma_{\mu\nu}\theta_\pm=\mp\varepsilon_{\mu\nu\rho}\Gamma^{\rho}\theta_\pm
\end{equation}
and
\begin{equation}\label{D1b}
\Gamma_{\hmu}\Gamma_{\el}\T_{\pm}=
\pm\varepsilon_{\hmu\hnu}\Gamma^{\hnu}\T_{\pm}
,\qquad
\Gamma_{\hmu}\tau_{1}\theta_\pm=\pm\varepsilon_{\hmu\hnu}
\Gamma^{\hnu}\theta_\pm
\label{gami2}
\end{equation}
for the F$1$ and D$1$ strings.
 
The differential forms and the spinors have independent gradings.
For $A^{p,r}$,$B^{q,s}$ of the form grades $p,q$ and the Grassman parities
$r,s$
\be
A^{p,r}B^{q,s}=(-1)^{pq+rs}\;B^{q,s}A^{p,r}.
\ee
Components of the forms are
defined by
\begin{equation} A_{r}=\frac{1}{r!}A_{i_1\ldots
i_r}\rmd\xi^{i_1}\ldots \rmd\xi^{i_r},
\end{equation}
and differentials are taken from the left.

\section{DBI action from  D$2$-brane Polyakov action}\label{apppoltong}

We have seen that the action of D$2$-brane in the Polyakov form
is obtained by dualizing $X^{\el}$ of the M$2$-brane action, \bref{D2Poly}. 
The kinetic term is
 \bea
S^{Pol}&=&-{T_2}\int d^3\xi\frac{ \sqrt{-\gamma}}{2}
\left\{\gamma^{IJ}G^{D2}_{IJ}-1
+\frac12\gamma^{IK}\gamma^{JL}\CF_{IJ}\CF_{KL}
\right\}
\label{D2PolyA}\eea
while the WZ term does not depends on the metric $\gamma_{IJ}$.
Here we give a proof 
that it gives the DBI action \bref{D2DBI}
\bea
S^{BI}&=&-{T_2}\int d^3\xi\sqrt{-\det (G^{D2}_{IJ}+\CF_{IJ})}
\label{DBID2A}\eea
by integrating the worldvolume metric.
In order to show it we should solve the equation of motion of 
the worldvolume metric and 
put it back to the action to eliminate its dependence.

The  equation of motion by taking variation of the action 
\bref{D2PolyA} with respect to $\gamma_{IJ}$, is
\be
\gamma_{IJ}=\frac2{\CX-\CY-1}\left(G^{D2}_{IJ}-\CF_{IK}\gamma^{KL}\CF_{LJ}
\right),
\label{gam1}
\ee
where we have defined
\be
\CX=\gamma^{IJ}G^{D2}_{IJ},\qquad \CY=\frac12
\gamma^{IJ}\CF_{JK}\gamma^{KL}\CF_{LI}.
\label{XYdef}
\ee
Since we are in $3$ worldvolume dimensions it is convenient to introduce
independent three vector $\CB^K$ by
\be
\CF_{IJ}=\varepsilon_{IJK}\CB^{K}
\ee
so that \bref{gam1} is
\be
\gamma_{IJ}=\frac2{\CX-\CY-1}\left(G^{D2}_{IJ}
+\frac{1}{\det\gam}\{\gamma_{IJ}(\CB^K\gamma_{KL}\CB^L)-(\gamma_{IK}\CB^K)
(\gamma_{JL}\CB^L)\}\right),
\label{gam2}
\ee
and $\CY$ in \bref{XYdef} is
\be
\CY=-\frac{\left(\CB^I\gamma_{IJ}\CB^J\right)}{\det{\gamma}}\equiv 
-\frac{(\CB \gamma \CB)}{{\gamma}}
\label{eq1}\ee
Multiplying $\gamma^{IJ}$ on \bref{gam2}
\be
3=\frac2{\CX-\CY-1}\left(\CX
+\frac{2}{\gam}(\CB\gamma\CB)\right).
\label{eq2}
\ee
Multiplying $\CB^I\CB^J$ on \bref{gam2}
\be
(\CB\gamma\CB)=\frac2{\CX-\CY-1}\left(\CB G^{D2}\CB\right).
\label{eq3}\ee
Solving \bref{eq1} and \bref{eq2}  for $\CX$ and $(\CB\gamma\CB)$
as
\be
\CX=3-\CY,\qquad (\CB\gamma\CB)=-\gamma\;\CY
\ee
and using them in \bref{eq3} 
\be
\gamma\;\CY(1-\CY)+(\CB G^{D2}\CB)=0.
\label{eq4}\ee
Return them back to  \bref{gam2} 
\be
\gamma_{IJ}=G^{D2}_{IJ}-\frac1{\gamma\;(1-\CY)^2}
(G^{D2}_{IK}\CB^K)
(G^{D2}_{JL}\CB^L),
\label{gam3}
\ee
and computing $\gamma=(\det\gamma)$ we get
\be
\gamma= G^{D2}-\frac{G^{D2}(\CB G^{D2}\CB)}{
\gamma\;(1-\CY)^2}=0.
\label{eq5}\ee
\bref{eq4} and \bref{eq5} are solved as
\be
\CY\=-\frac{(\CB G^{D2}\CB)}{G^{D2}},\qquad
\gamma\=\frac{ (G^{D2})^2}{(\CB G^{D2}\CB)+G^{D2}}
\ee  
and we determine $\gamma_{IJ}$ in terms of $G^{D2}_{IJ}$ and $\CB^I$ as
\be
\gamma_{IJ}=G^{D2}_{IJ}-\frac{(G^{D2}_{IK}\CB^K)
(G^{D2}_{JL}\CB^L)}{(\CB G^{D2}\CB)+G^{D2}}.
\label{gam5}
\ee
Finally we use it in the Lagrangian density of \bref{D2PolyA}. 
By squaring it becomes
\be
\cL^2=\frac{-\gamma}{4}(\CX-1-\CY)\=-
\left(G^{D2}+(\CB G^{D2}\CB)\right)\=
-\det(G^{D2}_{IJ}+\CF_{IJ}).
\ee
and the DBI action  \bref{DBID2A} follows from it.


\section{$\w^{-2}$ order term of $\gamma_{IJ}$ 
does not contribute}\label{appprof}

As we have seen in section 3 divergent term of $\cL^P$ \bref{pdivL} is
\bea \label{pdivLA}
\cL_{\textrm{div}}^P=-\frac{\sqrt{-\tgamma}}2\left(\tgamma^{IJ}g_{IJ}-1\right),
\eea
and on the other hand the $\tgamma_{ij}$ from the Lagrange equation of 
\bref{LagD2Poly} is 
\be
\tgamma_{IJ}=g_{IJ}+\frac1{\w^2}T_{IJ}+\mathcal{O}\left(\w^{-4}\right)
\label{tgammaexp}
\ee
A priori, we can expect that the $\w^{-2}$ order term of this expansion 
will give  finite contribution in the divergent $\w^2\cL_{\textrm{div}}^P$.
However it is easy to show that this is not what happens. 
Introducing the expansion of $\tgamma_{IJ}$ in $\cL^P_{\textrm{div}}$
\bea
\cL^P_{\textrm{div}}&=&-\frac{\sqrt{-g}}2
\left(1+\frac1{\w^2}g^{IJ}T_{IJ}\right)^{\frac12}
\left((g^{IJ}-\frac1{\w^2}T^{IJ})g_{IJ}-1\right)\nn\\
&=&-\frac{\sqrt{-g}}2\left(1+\frac1{2\w^2}g^{IJ}T_{IJ}\right)\left(2-\frac1{\w^2}T^{IJ}g_{IJ}\right)\nn\\
&=&-\sqrt{-g}+\mathcal{O}\left(\w^{-4}\right)
\eea
where $T^{IJ}=g^{IK}T_{KL}g^{LJ}$. 
It shows that $\w^{-2}$ order term of $\tgamma_{IJ}$ in \bref{tgammaexp}
does not give finite contribution in the divergent Lagrangian
$\w^2\cL^P_{\textrm{div}}$.


\end{appendix}


\begin{thebibliography}{99}

\bibitem{Gomis:2000bd}
J.~Gomis and H.~Ooguri,
``Non-relativistic closed string theory,''
J.\ Math.\ Phys.\  {\bf 42} (2001) 3127
[arXiv:hep-th/0009181].

\bibitem{Danielsson:2000gi}
U.~H.~Danielsson, A.~Guijosa and M.~Kruczenski,
``IIA/B, wound and wrapped,''
JHEP {\bf 0010} (2000) 020
[arXiv:hep-th/0009182].

\bibitem{Brugues:2004an}
J.~Brugues, T.~Curtright, J.~Gomis and L.~Mezincescu,
``Non-relativistic strings and branes as non-linear realizations of Galilei
groups,''
arXiv:hep-th/0404175.

\bibitem{Gomis:2004pw}
  J.~Gomis, K.~Kamimura and P.~K.~Townsend,
  ``Non-relativistic superbranes,''
  JHEP {\bf 0411}, 051 (2004)
  [arXiv:hep-th/0409219].


\bibitem{Gomis:2005pg}
  J.~Gomis, J.~Gomis and K.~Kamimura,
  ``Non-relativistic superstrings: A new soluble sector of AdS(5) x S**5,''
  arXiv:hep-th/0507036.
  
\bibitem{Gomis:2005bj}
  J.~Gomis, F.~Passerini, T.~Ramirez and A.~Van Proeyen,
  ``Non relativistic Dp branes,''
  arXiv:hep-th/0507135.
  
\bibitem{Duff:1987bx}
  M.~J.~Duff, P.~S.~Howe, T.~Inami and K.~S.~Stelle,
  ``Superstrings In D = 10 From Supermembranes In D = 11,''
  Phys.\ Lett.\ B {\bf 191} (1987) 70.

\bibitem{Townsend:1995af}
  P.~K.~Townsend,
  ``D-branes from M-branes,''
  Phys.\ Lett.\ B {\bf 373} (1996) 68
  [arXiv:hep-th/9512062].

\bibitem{Bergshoeff:1996cy}
  E.~Bergshoeff and M.~De Roo,
  ``D-branes and T-duality,''
  Phys.\ Lett.\ B {\bf 380} (1996) 265
  [arXiv:hep-th/9603123].


\bibitem{Green:1996bh}
M.~B.~Green, C.~M.~Hull and P.~K.~Townsend,
``D-Brane Wess-Zumino Actions, T-Duality and the Cosmological Constant,''
Phys.\ Lett.\ B {\bf 382}, 65 (1996)
[arXiv:hep-th/9604119].


\bibitem{Bergshoeff:1987cm}
E.~Bergshoeff, E.~Sezgin and P.~K.~Townsend,
``Supermembranes And Eleven-Dimensional Supergravity,''
Phys.\ Lett.\ B {\bf 189}, 75 (1987).

\bibitem{Cederwall:1996pv}
M.~Cederwall, A.~von Gussich, B.~E.~W.~Nilsson and A.~Westerberg,
``The Dirichlet super-three-brane in ten-dimensional type IIB  supergravity,''
Nucl.\ Phys.\ B {\bf 490}, 163 (1997)
[arXiv:hep-th/9610148].

\bibitem{Aganagic:1996pe}
M.~Aganagic, C.~Popescu and J.~H.~Schwarz,
``D-brane actions with local kappa symmetry,''
B {\bf 393}, 311 (1997)
[arXiv:hep-th/9610249].


\bibitem{Cederwall:1996ri}
M.~Cederwall, A.~von Gussich, B.~E.~W.~Nilsson, P.~Sundell and A.~Westerberg,
``The Dirichlet super-p-branes in ten-dimensional type IIA and IIB
supergravity,''
Nucl.\ Phys.\ B {\bf 490}, 179 (1997)
[arXiv:hep-th/9611159].

\bibitem{Bergshoeff:1996tu}
E.~Bergshoeff and P.~K.~Townsend,
``Super D-branes,''
Nucl.\ Phys.\ B {\bf 490}, 145 (1997)
[arXiv:hep-th/9611173].

\bibitem{Green:1983wt}
M.~B.~Green and J.~H.~Schwarz,
``Covariant Description Of Superstrings,''
Phys.\ Lett.\ B {\bf 136} (1984) 367.

\bibitem{Simon:2000bi}
J.~Simon,
``T-duality and Effective D-Brane Actions,''
Phys.\ Rev.\ D {\bf 61} (2000) 047702
[arXiv:hep-th/9812095].


\bibitem{Kamimura:2000bi}
K.~Kamimura and J.~Simon,
``T-duality covariance of superD-branes,''
Nucl.\ Phys.\ B {\bf 585}, 219 (2000)
[arXiv:hep-th/0003211].



\end{thebibliography}
\end{document}